\documentclass[aps, prl, twocolumn, superscriptaddress,longbibliography,nobibnotes]{revtex4-1}
\usepackage{graphicx}
\usepackage[hidelinks]{hyperref} 
\usepackage{bm}
\usepackage{amsmath}
\usepackage{bbold}

\graphicspath{./}

\begin{document}


\title{Spiral spin-liquid on a honeycomb lattice} 
\thanks{This manuscript has been authored by UT-Battelle, LLC under Contract No. DE-AC05-00OR22725 with the U.S. Department of Energy.  The United States Government retains and the publisher, by accepting the article for publication, acknowledges that the United States Government retains a non-exclusive, paid-up, irrevocable, world-wide license to publish or reproduce the published form of this manuscript, or allow others to do so, for United States Government purposes.  The Department of Energy will provide public access to these results of federally sponsored research in accordance with the DOE Public Access Plan (http://energy.gov/downloads/doe-public-access-plan).}

\renewcommand*{\thefootnote}{\arabic{footnote}}

\author{Shang Gao}
\email[]{sgao.physics@gmail.com}
\affiliation{Neutron Scattering Division, Oak Ridge National Laboratory, Oak Ridge, TN 37831, USA}
\affiliation{Materials Science \& Technology Division, Oak Ridge National Laboratory, Oak Ridge, TN 37831, USA}

\author{Michael A. McGuire}
\affiliation{Materials Science \& Technology Division, Oak Ridge National Laboratory, Oak Ridge, TN 37831, USA}

\author{Yaohua Liu}
\affiliation{Neutron Scattering Division, Oak Ridge National Laboratory, Oak Ridge, TN 37831, USA}

\author{Douglas L. Abernathy}
\affiliation{Neutron Scattering Division, Oak Ridge National Laboratory, Oak Ridge, TN 37831, USA}

\author{Clarina dela Cruz}
\affiliation{Neutron Scattering Division, Oak Ridge National Laboratory, Oak Ridge, TN 37831, USA}

\author{Matthias Frontzek}
\affiliation{Neutron Scattering Division, Oak Ridge National Laboratory, Oak Ridge, TN 37831, USA}

\author{Matthew B. Stone}
\affiliation{Neutron Scattering Division, Oak Ridge National Laboratory, Oak Ridge, TN 37831, USA}

\author{Andrew D. Christianson}
\affiliation{Materials Science \& Technology Division, Oak Ridge National Laboratory, Oak Ridge, TN 37831, USA}

\date{\today}


\begin{abstract}
  Spiral spin-liquids are correlated paramagnetic states with degenerate propagation vectors forming a continuous ring or surface in reciprocal space. On the honeycomb lattice, spiral spin-liquids present a novel route to realize emergent fracton excitations, quantum spin liquids, and topological spin textures, yet experimental realizations remain elusive. Here, using neutron scattering, we show that a spiral spin-liquid is realized in the van der Waals honeycomb magnet FeCl$_3$. A continuous ring of scattering is directly observed, which indicates the emergence of an approximate U(1) symmetry in momentum space. Our work demonstrates that spiral spin-liquids can be achieved in two-dimensional systems and provides a promising platform to study the fracton physics in spiral spin-liquids.
  \end{abstract}
  
  \maketitle
  
  Similar to geometrical frustration~\cite{bramwell_spin_2001, balents_spin_2010}, competition amongst interactions at different length scales is able to induce novel  electronic or magnetic states regardless of the lattice geometry. A representative example is the spiral spin-liquid (SSL), which is a type of classical spin liquid realized on a bipartite lattice~\cite{bergman_order_2007, lee_theory_2008, mulder_spiral_2010, chen_frustrated_2012, zhang_exotic_2013, niggemann_classical_2019, pohle_theory_2021, attig_classical_2017, iqbal_stability_2018, balla_degenerate_2020, balla_affine_2019, lee_theory_2008, chen_quantum_2017,shimokawa_ripple_2019, yao_generic_2021, huang_versatility_2021, buessen_quantum_2018, liu_featureless_2020, yan_low_2021}. In such a state, spins fluctuate collectively as spirals, and their propagation vectors, $\bm{q}$, form a continuous ring or surface in reciprocal space. Depending on the specific shape of the spiral surface, the low-energy fluctuations in a SSL may behave as topological vortices in momentum space, leading to an effective tensor gauge theory with highly unconventional fracton quadrupole excitations~\cite{yan_low_2021, pretko_sub_2017, nandkishore_fractons_2019, pretko_fracton_2020}. Such nonlocal dynamics is very different from that in geometrically frustrated magnets, where the elementary excitations are local spin flips. Compared to the conventional frustrated geometry, a bipartite lattice offers more flexibility on the signs of the interactions, as the duality between antiferromagnetic and ferromagnetic interactions indicates that a SSL can be realized even in ferromagnets as long as sufficient competition exists~\cite{bergman_order_2007}. Since degeneracy enhances quantum fluctuations~\cite{balents_spin_2010,zhou_quantum_2017}, the SSL has thus been proposed as a novel route to realize quantum spin liquids in systems dominated by ferromagnetic interactions~\cite{mulder_spiral_2010, chen_frustrated_2012, zhang_exotic_2013, niggemann_classical_2019,pohle_theory_2021}. Furthermore, when perturbations, \textit{e.g.} the further-neighbor interactions or anisotropic interactions, induce a magnetic long-range order, degeneracy in the SSL may be partially retained, leading to skyrmion-like topological spin textures~\cite{gao_spiral_2017, gao_fractional_2020, shimokawa_multiple_2019} that have great potential for applications in spintronic devices~\cite{nagaosa_topological_2013,fert_magnetic_2017}.
  
  
  \begin{figure}[b!]
      \includegraphics[width=0.48\textwidth]{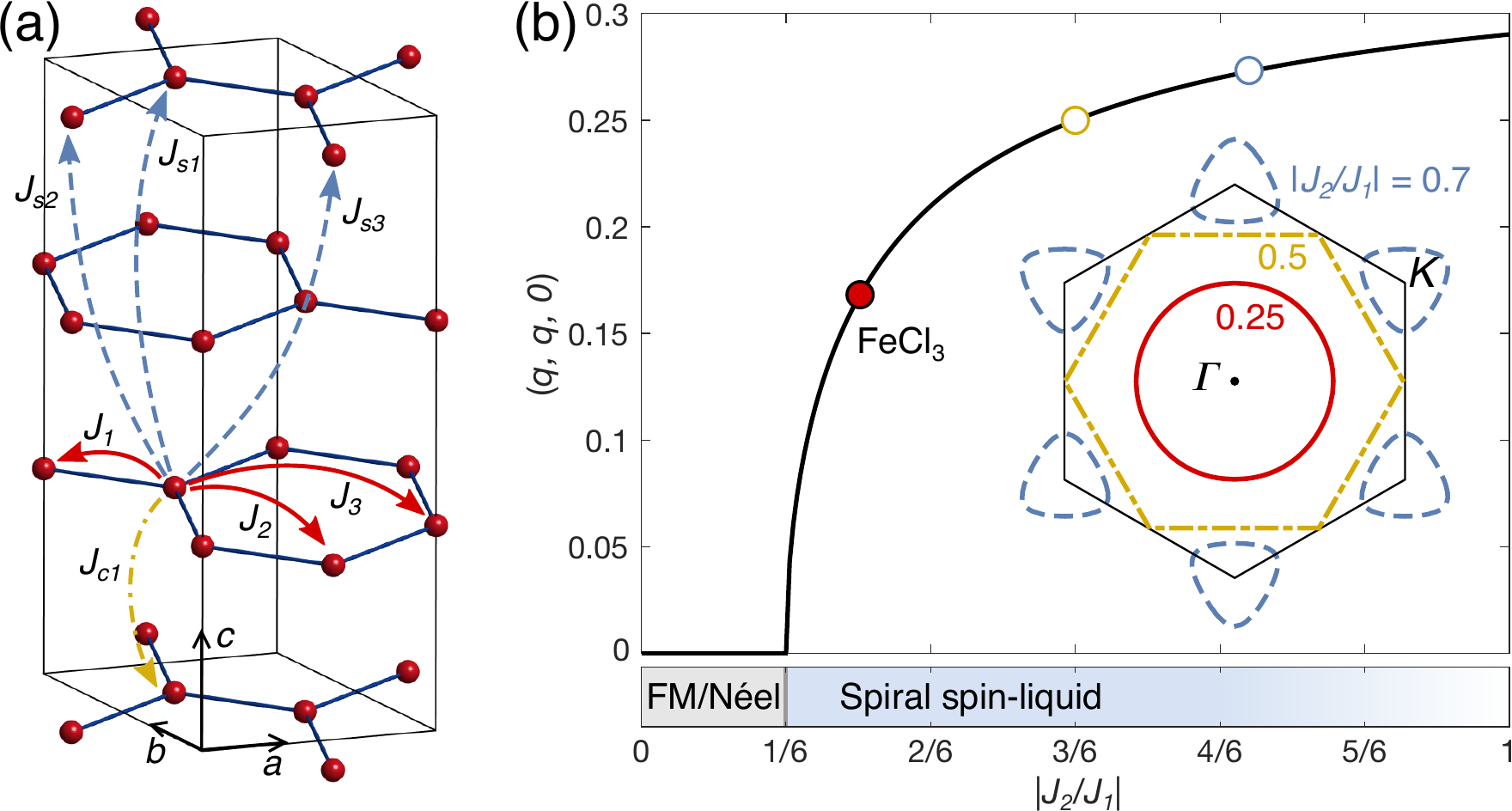}
      \caption{(a) The Fe$^{3+}$ ions ($S=5/2$) in FeCl$_3$ form honeycomb lattices with $ABC$-type stacking along the $c$ axis. Red solid arrows indicate the nearest-, second-, and third-neighbor couplings $J_1$, $J_2$, and $J_3$, respectively. Yellow dot-dashed arrow indicates the interlayer couplings $J_{c1}$. Blue dashed arrows indicate the second-layer couplings $J_{s1}$, $J_{s2}$, and $J_{s3}$. (b) A spiral spin-liquid state is realized on the honeycomb lattice at $|J_2/J_1|> 1/6$ (blue shaded in the bottom panel) with propagation vectors forming a continuous ring in reciprocal space, which we refer to as the spiral ring. The black curve in the top panel shows the position of a representative propagation vector ($q$, $q$, 0) over the spiral ring as a function of $|J_2/J_1|$. Inset shows the complete spiral rings at $|J_2/J_1| = 0.25$, 0.5, and 0.7 as indicated by circular markers over the black curve. The red filled marker indicates the location of FeCl$_3$ with an effective ratio of $|J_2/J_1|\sim0.25$ as determined from the $J_1$-$J_2$-$J_c$ minimal model (see text).
      \label{fig:intro}}
      \end{figure}
  
  On a bipartite lattice with Heisenberg interactions, a SSL emerges when the ratio between the effective second-neighbor and first-neighbor couplings $|J_2^*/J_1|$ is higher than a threshold of $1/(2Z)$, where $Z$ counts the number of the nearest-neighboring sites~\cite{bergman_order_2007,niggemann_classical_2019}. Depending on the exact lattice geometry, $J_2^*$ may include contributions from further-neighbor interactions at fixed ratios~\cite{niggemann_classical_2019,balla_affine_2019,yan_low_2021}. However, on the honeycomb or diamond lattice where the sum of two successive $J_1$ paths always equals a $J_2$ path~\cite{mulder_spiral_2010, bergman_order_2007}, further-neighbor contributions to $J_2^*$ are not necessary to stabilize a SSL ($J_2^*=J_2$), which greatly simplifies the experimental realization. The phase diagram of a $J_1$-$J_2$ model on a honeycomb lattice~\cite{mulder_spiral_2010} is summarized in Fig.~\ref{fig:intro}, where the evolution of the representative $\bm{q}=(q, q, 0)$ over the degenerate ground state manifold is shown explicitly. Following the definition of the spiral surface on the diamond lattice~\cite{bergman_order_2007}, we call the degenerate ring in the SSL state a \textit{spiral ring}. Above the threshold of $|J_2/J_1|=1/6$, the ferromagnetic (FM, $J_1<0$) or N\'eel ($J_1>0$) state with $q$ = 0 is replaced by a SSL state with nonzero $q$, and the spiral ring gradually transforms from a circular shape around the Brillouin zone center, $\Gamma$, into triangular lobes centered on the $K~\{\frac{1}{3},\frac{1}{3}\}$ points as $|J_2/J_1|$ increases. The case with a spiral ring around $\Gamma$ is extremely interesting, as the low-energy dynamics can be described as fracton quadrupoles in a rank-2 U(1) tensor gauge theory~\cite{yan_low_2021}, which is a heavily investigated field with deep connections to quantum information, elasticity, and gravity~\cite{pretko_emergent_2017, halasz_fracton_2017, pretko_fracton_2018, yan_hyperbolic_2019, nandkishore_fractons_2019,pretko_fracton_2020}.
  
  In spite of the elegant simplicity of the theoretical model, experimental realization of a SSL is challenging as $J_2$ is often relatively weak in real materials~\cite{supp, gao_spiral_2017, macdougall_revisiting_2016, ge_spin_2017, chamorro_frustrated_2018, tsurkan_on_2021, haraguchi_frustrated_2019, abdeldaim_realizing_2020, otsuka_canting_2020}. To our knowledge, MnSc$_2$S$_4$ has remained as the only host of a SSL on the diamond lattice~\cite{gao_spiral_2017}, while the feasibility of realizing a SSL on the honeycomb lattice is still unclear. Here we show that a SSL state with an approximate U(1) symmetry in momentum space is realized in the honeycomb magnet FeCl$_3$~\cite{cable_neutron_1962, jones_low_1969, stampfel_mossbauer_1973, johnson_field_1981, Hashimoto_structure_1989, kang_magnetic_2014}. This compound belongs to the van der Waals trihalide family that has recently attracted great attention for its fundamental and application interests~\cite{ huang_layer_2017, banerjee_neutron_2017, mcguire_crystal_2017}. As shown in Fig.~\ref{fig:intro}(a), the honeycomb Fe$^{3+}$ ($S = 5/2$) layers in FeCl$_3$ are $ABC$-stacked along the $c$ axis, leading to a rhombohedral $R\overline{3}$ space group. Previous neutron diffraction experiments performed in the 1960s~\cite{cable_neutron_1962, jones_low_1969} revealed a helical magnetic long-range order (LRO) with $\bm{q}=(\frac{4}{15},\frac{1}{15},\frac{3}{2})$ below the transition temperature $T_N\sim 10$~K, which indicates an antiferromagnetic interlayer alignment and possible intralayer frustration. In the present paper, we utilize state-of-the-art neutron scattering measurements to study the short-range spin correlations above $T_N$. A continuous ring of scattering around $\Gamma$ is observed, which provides direct evidence for the existence of a SSL state with an approximate U(1) symmetry in momentum space. The spiral correlations can be mainly ascribed to the $J_1$-$J_2$ competition, which is further corroborated through inelastic neutron scattering (INS) in the long-range ordered phase. Details for sample preparations and neutron scattering experiments are presented in the Supplemental Materials~\cite{supp}.
  
  \begin{figure}[t!]
      \includegraphics[width=0.48\textwidth]{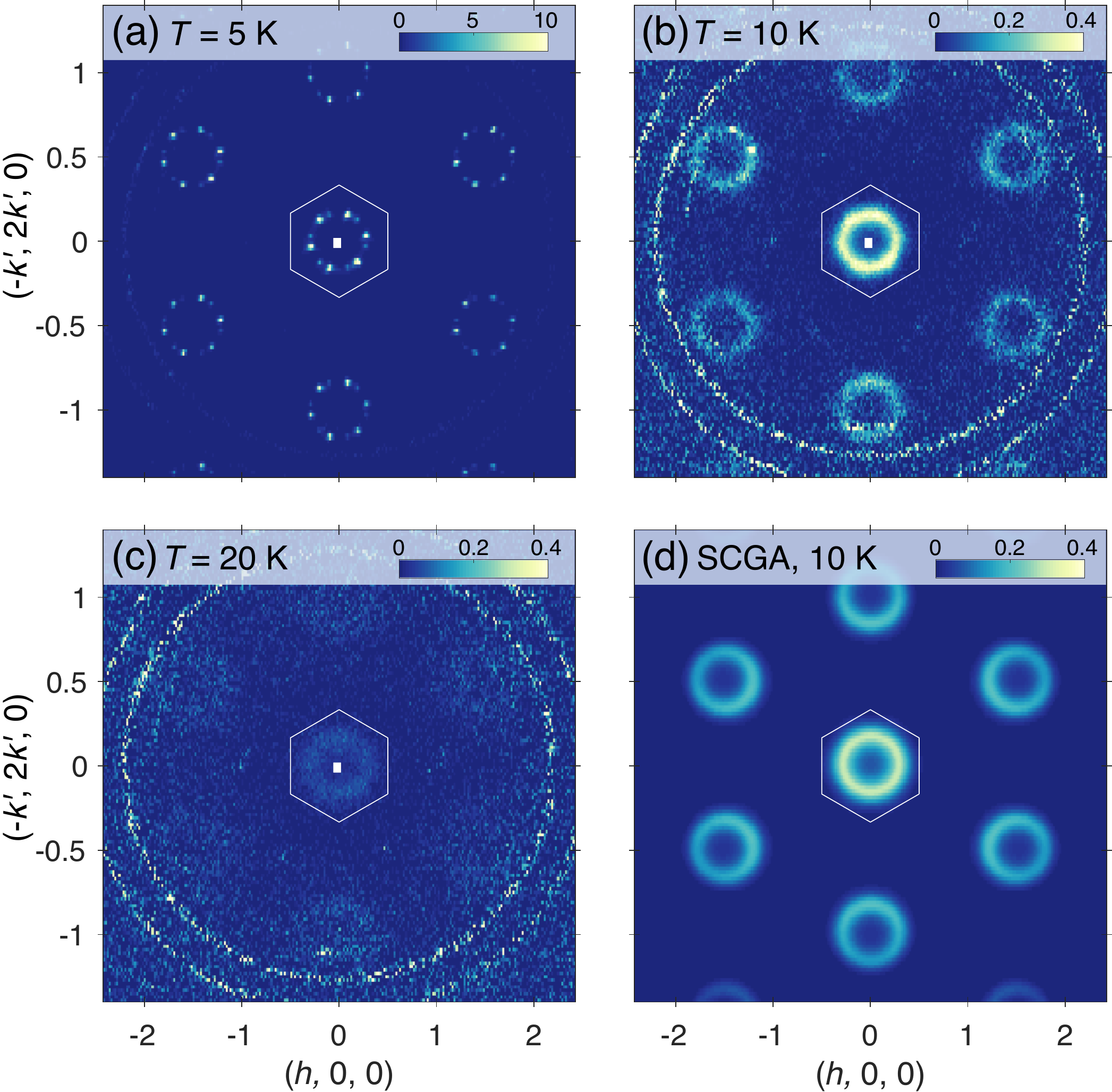}
      \caption{(a-c) Temperature evolution of the quasi-elastic spin correlations in the $l=-1.5$ plane measured on CORELLI at $T = 5$, 10, and 20~K. The white hexagon outlines the first nuclear Brillouin zone. Data are integrated in the range of $l=[-1.6, -1.4]$. Measurements at 50~K have been subtracted as the background. In (b) and (c), the double rings spanning the whole panel are background scattering from the sample environment, which is less evident in (a) due to the different intensity scale.  (d) Calculated diffuse neutron scattering pattern in the $l=-1.5$ plane using the minimal $J_1$-$J_2$-$J_{c1}$ model at $T=10$~K. The coupling strengths are $J_1 = -0.3$~meV, $J_2 = 0.075$~meV, and $J_{c1}=0.15$~meV. Variations in coupling strengths do not qualitatively affect the diffuse pattern as long as $J_2/J_1=-0.25$ with $J_1$ and $J_{c1}$ being ferromagnetic and antiferromagnetic, respectively.  
      \label{fig:tdep}}
      \end{figure}
  
  Diffuse neutron scattering probes the short-range spin correlations in reciprocal space. Figures \ref{fig:tdep}(a-c) summarize the temperature dependence of our diffuse scattering pattern in the $l = -1.5$ plane, as it exhibits the strongest intensity throughout reciprocal space. Below the transition temperature of $T_N \sim 8$~K, magnetic Bragg peaks belonging to $\bm{q}=(\frac{4}{15},\,\frac{1}{15},\frac{3}{2})$ are observed, which is consistent with the previous diffraction study~\cite{cable_neutron_1962}. At 10~K above $T_N$, the magnetic Bragg peaks merge together, leading to a ring of scattering that implies an emergent U(1) symmetry in momentum space~\cite{yan_low_2021}. This scattering ring can be discerned at temperatures up to $\sim20$~K as shown in Fig.~\ref{fig:tdep}(c), indicating a relatively wide stability regime for the spiral correlations.
  
  \begin{figure*}[t!]
      \includegraphics[width=1.0\textwidth]{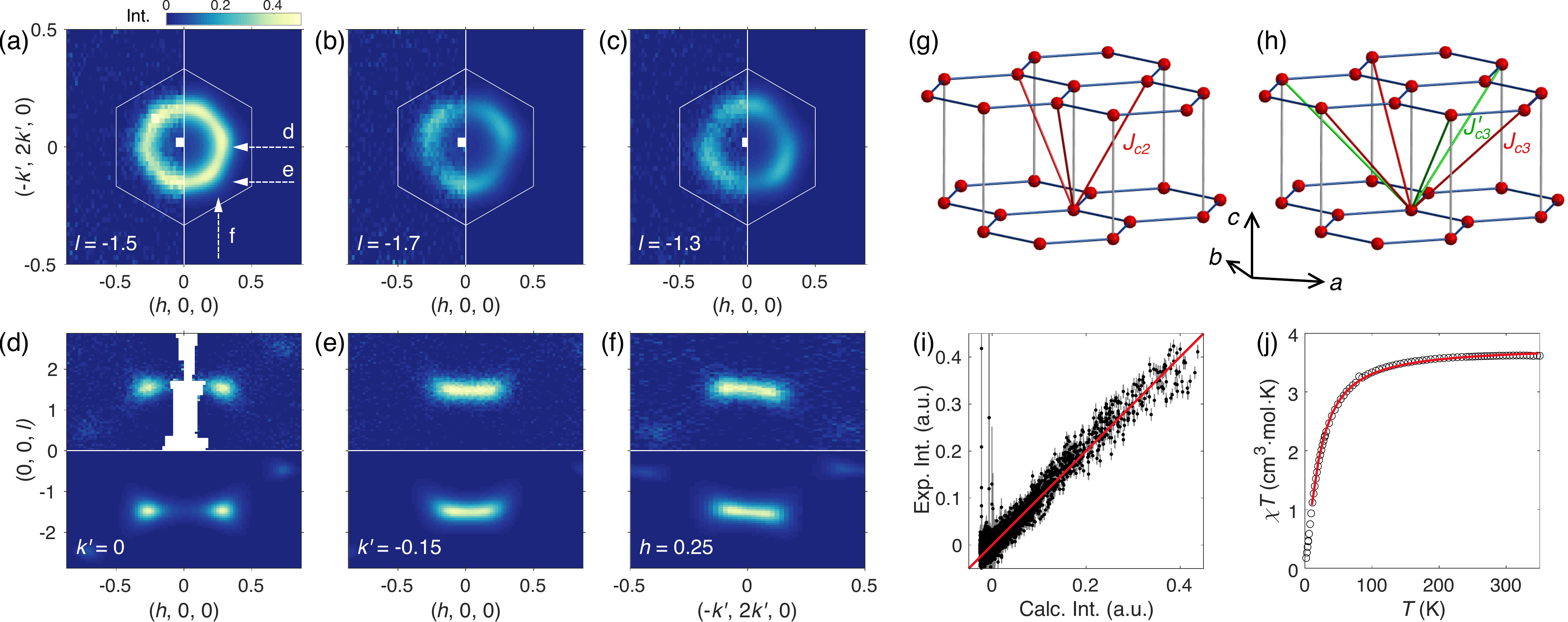}
      \caption{(a-f)
      Diffuse neutron scattering intensity measured at 10 K together with the simulated results. Data in the $l = -1.5$, $-1.7$, and $-1.3$ planes are shown in the left half of panels (a-c), respectively. Data in the $k'=0$, $k'=-0.15$, and $h = 0.25$ planes are shown in the upper half of panels (d-f), respectively. In panels (a-c), the white hexagon outlines the first nuclear Brillouin zone. Dashed arrows in panel (a) indicate the directions of the vertical slices in panels (d-f). (g,h) Exchange paths for the interlayer interactions (g) $J_{c2}$ (red solid line) and (h) $J_{c3}$ (red solid line). In panel (h), the $J'_{c3}$ interactions (green solid line) are symmetrically inequivalent with the $J_{c3}$ interactions in spite of their equal distances. (i) Comparison of the experimental and calculated diffuse scattering intensities using the $J_{123}$-$J_{c123}$ model. The fitted coupling strengths are $J_1 = -0.249(4)$~meV, $J_2=0.089(1)$~meV, $J_3=0.026(1)$~meV, $J_{c1}=0.019(9)$~meV, $J_{c2}=0.042(2)$~meV, $J_{c3}=0.030(2)$~meV. Uncertainties are estimated from 50 independent runs.  The goodness-of-fit factor is $\chi^2=1.77$.  (j) Temperature dependence of the reduced magnetic susceptibility $\chi T$ (black circle) together with the calculated values based upon the SCGA method described in the text (red line). Data are measured on a powder sample in a 1~T magnetic field~\cite{supp}. Error bars representing the standard deviations are smaller than the symbols. 
      \label{fig:scga}}
      \end{figure*}
  
  To confirm that the diffuse ring of scattering originates from a SSL state, we calculate the short-range spin correlations using the self-consistent Gaussian approximation (SCGA) method~\cite{supp}. The fact that the diffuse scattering intensity at 10~K is concentrated in the half integer $l$ planes suggests antiferromagnetic correlations between the neighboring honeycomb layers. Therefore, in addition to the $J_1$ and $J_2$ interactions within the honeycomb layers, we consider an antiferromagnetic interlayer interaction $J_{c1}$ along the $c$ axis as shown in Fig.~\ref{fig:intro}(a). With ferromagnetic $J_1$ and a frustration ratio of $J_2/J_1 = -0.25$, the calculated pattern presented in Fig.~\ref{fig:tdep}(d) captures the scattering ring observed at $T = 10$~K, thus establishing the existence of an intrinsic SSL state in FeCl$_3$. The effective ratio of $|J_2/J_1| = 0.25$ also grants a good approximation of the U(1) symmetry in reciprocal space since higher ratios may introduce a strong distortion of the circular shape as compared in Fig.~\ref{fig:intro}(b).
  
  Although the $J_1$-$J_2$-$J_{c1}$ minimal model successfully reproduces the spiral ring in the $l=-1.5$ plane, it is too simplified to describe the full spin correlations in FeCl$_3$. Figures~\ref{fig:scga}(a-f) summarize the detailed intensity distribution of the spiral ring. The corresponding data in a wider range are shown in the Supplemental Materials~\cite{supp}. The scattering intensity on the two sides of the $l=-1.5$ plane exhibits reversed three-fold symmetry patterns, which is not reproduced by the $J_1$-$J_2$-$J_{c1}$ model~\cite{supp} and suggests a weak U(1) symmetry breaking due to further perturbations. 
  \begin{figure*}[t!]
      \includegraphics[width=1.0\textwidth]{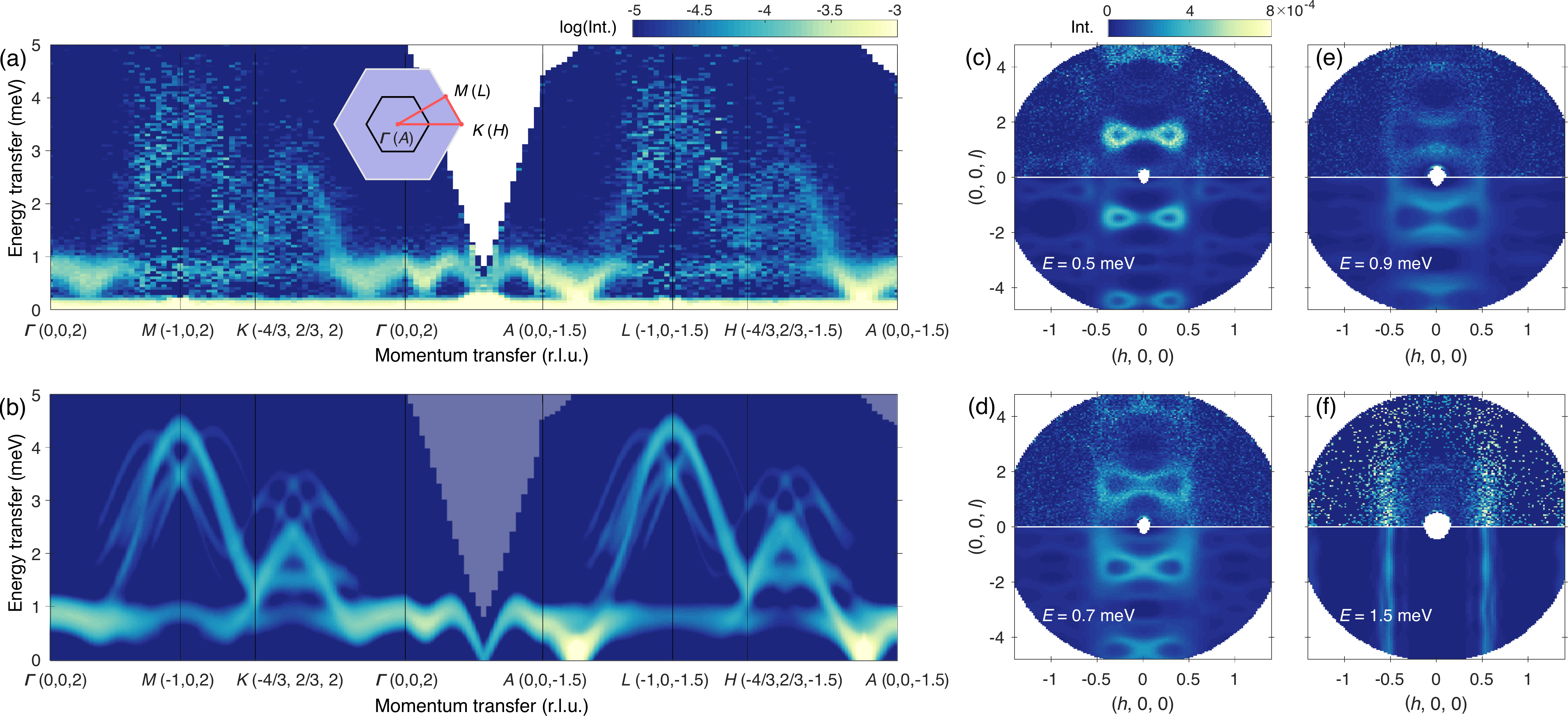}
      \caption{(a) Experimental INS spectra $S(\bm{Q}, \omega)$ measured on SEQUOIA with an incident neutron energy of $E_i=8$~meV  at $T = 4$~K along the high symmetry directions. Data are symmetrized according to the $R\overline{3}$ symmetry. The positions of the high symmetry points in the $l=2$ and $l=-1.5$ planes are shown in the inset, with the first nuclear Brillouin zone being outlined by the black hexagon. Intensity is plotted in a log scale. (b) Calculated INS spectra using the linear spin wave theory for a $J_{123}$-$J_{c123}$-$J_{s123}$ model with fitted coupling strengths of $J_1 = -0.28(3)$~meV, $J_2 = 0.095(9)$~meV, $J_3 = 0.008(5)$~meV, $J_{c1}=0.05(3)$~meV, $J_{c2}=0.024(5)$~meV, $J_{c3}=0.026(6)$~meV, $J_{s1}=0.01(1)$~meV, $J_{s2}=-0.007(3)$~meV, and $J_{s3}=-0.003(4)$~meV. Uncertainties are estimated from 20 independent runs. A weak easy plane single-ion anisotropy of $1~{\mu}$eV is included to stabilize the helical ground state. The calculated spectra are convoluted by a Gaussian function with a fitted full-width at half-maximum (FWHM) of 0.35 meV. Intensity is plotted in a log scale as that for the experimental data. (c-f) Comparison between the experimental and calculated constant-energy slices at $E=0.5$ (c), 0.7 (d), 0.9 (e), and 1.5 (f) meV. Data are integrated in an energy range of $\pm 0.1$~meV. Intensity in panel (f) is multiplied by 8 times for better visibility. The same linear intensity scale as shown in panel (c) is utilized for panels (c-f). 
      \label{fig:ins}}
  \end{figure*}
  
  Using the SCGA method, we explore the effects on the spiral ring from perturbations that are allowed by the $R\overline{3}$ symmetry. As discussed in the Supplemental Materials~\cite{supp}, perturbations from the anisotropic interactions including the Kitaev-like interactions or the relativistic Dzyaloshinskii-Moriya (DM) interactions are not able to reproduce the diffuse scattering data. Therefore, we concentrate on the isotropic further-neighbor interactions that are consistent with the quenched orbital degree of freedom of the Fe$^{3+}$ ions. Besides the third-neighbor coupling $J_3$ within the honeycomb layer shown in Fig.~\ref{fig:intro}(a), two additional interlayer couplings $J_{c2}$ and $J_{c3}$ are found to be important in explaining the diffuse scattering data. Figures~\ref{fig:scga}(g) and (h) present the exchange paths for the $J_{c2}$ and $J_{c3}$ interactions. The $J_{c3}$ interaction couples the spin at the origin to those at $\bm{a}+\bm{c}/3$ and the equivalent positions, where $\bm{a}$ and $\bm{c}$ are the basis vectors of the hexagonal unit cell. This interaction should be differentiated from the equal-distant $J'_{c3}$ interaction shown by the green solid lines in Fig.~\ref{fig:scga}(h), as their exchange paths are subject to different symmetry constraints. Using the combined simplex/simulated annealing optimization method~\cite{farhi_ifit_2014}, we fit the diffuse scattering data over volumes of reciprocal space together with the temperature dependence of the magnetic susceptibility $\chi(T)$. The fitted results are presented in Figs.~\ref{fig:scga}(i) and (j) for the volume diffuse scattering data and the reduced magnetic susceptibility, respectively, and the fitted diffuse patterns are presented in Fig.~\ref{fig:scga}(a-f) together with the experimental results. The fitted coupling strengths as listed in the caption of Fig.~\ref{fig:scga} reveal a relatively high frustration ratio of $|J_2/J_1| = 0.36$, thus confirming the $J_1$-$J_2$ competition as the driving force of the SSL state in FeCl$_3$.
  
  Greater insight into the spin interactions emerges through the analysis of the spin excitations. Figure~\ref{fig:ins}(a) summarizes the INS spectra $S(\bm{Q}, \omega)$ of FeCl$_3$ measured at $T = 5$~K in the helical ordered phase. Magnon excitations emanating from the LRO $\bm{q}=(\frac{4}{15},\,\frac{1}{15},\,\frac{3}{2})$ are observed throughout reciprocal space. Two magnon branches can be discerned in the energy ranges of [0, 1.0] and  [1.0, 4.0]~meV, which can be correspondingly attributed to the in-phase and anti-phase movements of the two sublattice spins on the honeycomb lattice. The gapless excitations along the $A$-$L$-$H$-$A$ line in the $l=-1.5$ plane are consistent with the ground state degeneracy caused by the $J_1$-$J_2$ competition. Compared to the instrumental resolution of $\sim 0.19$~meV at the elastic line, the observed magnons, especially the [1.0, 4.0]~meV branch, exhibit a broader width, up to $\sim 1$~meV, indicating unresolved magnon modes due to multiple magnetic domains together with the separated $\bm{Q} \pm \bm{q}$ excitations~\cite{toth_linear_2015}. In contrast, all these modes overlap along the $c$ axis, which leads to a better-resolved gull wing-shaped dispersion along the $\Gamma$-$A$ line.
  
  To understand the spin excitations in FeCl$_3$, we perform linear spin wave calculations for the Heisenberg spin Hamiltonian. As discussed in the Supplemental Materials~\cite{supp}, the $J_{123}$-$J_{c123}$ model captures the main features of the magnon dispersion in the honeycomb layers but produces a dispersion along the $c$ axis that is clearly weaker than the observed spectrum. Therefore, three additional couplings $J_{s1}$, $J_{s2}$, and $J_{s3}$ are included in the spin Hamiltonian, which are the shortest exchange interactions between the second-neighboring layers as shown in Fig.~\ref{fig:intro}(a). By fitting the INS spectra at $E<1.0$~meV, we arrive at the parameter set listed in the caption of Fig.~\ref{fig:ins}. The strengths of the second-layer couplings are relatively weak as expected from their longer exchange paths, and the strengths of the remaining interlayer and intralayer couplings are close to those fitted from the $J_{123}$-$J_{s123}$ model. The calculated INS spectra in Fig.~\ref{fig:ins}(b) reproduce the experimental data, and the good agreement is further confirmed through the comparison of the experimental and calculated constant energy slices shown in Figs.~\ref{fig:ins}(c-f). Meanwhile, the calculated diffuse scattering patterns for the $J_{123}$-$J_{c123}$-$J_{s123}$ model stay almost unchanged~\cite{supp}, suggesting that short-range spin correlations are not sensitive to the weak second-layer interactions.
  
  The perturbations on the minimal $J_1$-$J_2$-$J_{c1}$ model also explain the selection of the LRO $\bm{q}$ position over the spiral ring. Using the Luttinger-Tisza method~\cite{niggemann_classical_2019}, the ground state of the $J_1$-$J_2$-$J_{c1}$ model can be calculated to be a spiral with $\bm{q} = (\frac{1}{6},\frac{1}{6},\frac{3}{2})$, while the LRO $\bm{q}$ of the perturbed $J_{123}$-$J_{c123}$-$J_{s123}$ model become $(0.194, 0.096, 1.5)$, which is closer to the experimentally observed value of $\bm{q}=(\frac{4}{15},\,\frac{1}{15},\frac{3}{2})$. Although fine tuning of the coupling strengths seems necessary to exactly reproduce the commensurate $\bm{q}$ position in FeCl$_3$, theoretical calculations of SSLs have proposed a lock-in mechanism where an incommensurate $\bm{q}$ becomes pinned to a nearby commensurate position due to weak single-ion anisotropy and crystal symmetry~\cite{lee_theory_2008, supp}. Such a lock-in mechanism may account for the commensurate $\bm{q}$ position in FeCl$_3$.
  
  
  Our experimental study on FeCl$_3$ demonstrates that SSLs can be realized in two-dimensional systems. Remarkably, the observed spiral ring around $\Gamma$ implies an emergent U(1) symmetry in momentum space and establishes FeCl$_3$ as a promising platform to study the fracton gauge theory. This prospect is further encouraged because FeCl$_3$ can be easily cleaved into monolayers, allowing for the elimination of the out-of-plane perturbations, $J_c$ and $J_s$. The discovery of a SSL in FeCl$_3$ also motivates further investigations of quantum spin liquids and topological spin textures on the honeycomb lattice. Current experimental endeavors on quantum spin liquids are mainly focused on the Kitaev approach~\cite{banerjee_neutron_2017}, while the approach via spiral spin-liquid phase has remained barely explored~\cite{gong_phase_2013,niggemann_classical_2019}. Knowing the origin of the relatively high ratio of $|J_2/J_1|$ in FeCl$_3$, e.g. through the \textit{ab initio} calculations, will help discover more SSL hosts on the honeycomb lattice and facilitate the tuning of SSL towards the quantum limit or spintronics applications~\cite{baltz_anti_2018,jungwirth_multiple_2018,lu_meron_2020,augustin_properties_2021}.
  
  
  \begin{acknowledgments}
  We acknowledge helpful discussions with Cheng-Long Zhang at RIKEN CEMS. This work was supported by the U.S. Department of Energy, Office of Science, Basic Energy Sciences, Materials Sciences and Engineering Division. This research used resources at the Spallation Neutron Source (SNS) and the High Flux Isotope Reactor (HFIR), both are DOE Office of Science User Facilities operated by the Oak Ridge National Laboratory (ORNL). 
  \end{acknowledgments}
  

  %

  \clearpage
  \newpage
  
  \renewcommand{\thefigure}{S\arabic{figure}}
  \renewcommand{\thetable}{S\arabic{table}}
  \renewcommand{\theequation}{S\arabic{equation}}

  \makeatletter
  \renewcommand*{\citenumfont}[1]{S#1}
  \renewcommand*{\bibnumfmt}[1]{[S#1]}
  \def\clearfmfn{\let\@FMN@list\@empty}  
  \makeatother
  \clearfmfn

  \setcounter{figure}{0} 
  \setcounter{table}{0}
  \setcounter{equation}{0} 
  
  \onecolumngrid
  \begin{center} {\bf \large Supplemental materials for `spiral spin-liquid on a honeycomb lattice'} \end{center}
  
  \vspace{0.5cm}
   
  \subsection{Sample preparation}
  Crystals of FeCl$_3$ were grown using the Bridgman technique \cite{stampfel_mossbauer_1973s, may_practical_2020}, using commercial FeCl$_3$ powder from Sigma-Aldrich (99.99\%) as the starting material. The material is very air sensitive, and was handled in a glovebox under a purified helium atmosphere and never exposed to air throughout the growth, sample preparation, and measurement processes. For the growth of the crystals used in the present study, 15\,g of FeCl$_3$ was sealed inside an evacuated silica ampoule with an inner diameter of 16\,mm, an outer diameter 19\,mm, and a taper to a point at one end. The ampoule was suspended in a vertical tube furnace so the lower (pointed) end was at 330\,$^{\circ}$C and the top was at about 350\,$^{\circ}$C. The sample was held stationary throughout the growth, which started with a 6 hour hold at the starting conditions before cooling at 1\,$^{\circ}$C/hr until the lower tip temperature reached 250\,$^{\circ}$C. The furnace was then turned off, and the sample was removed once the tip temperature had cooled to 150\,$^{\circ}$C.
  
  \begin{figure}[b!]
      \includegraphics[width=0.4\textwidth]{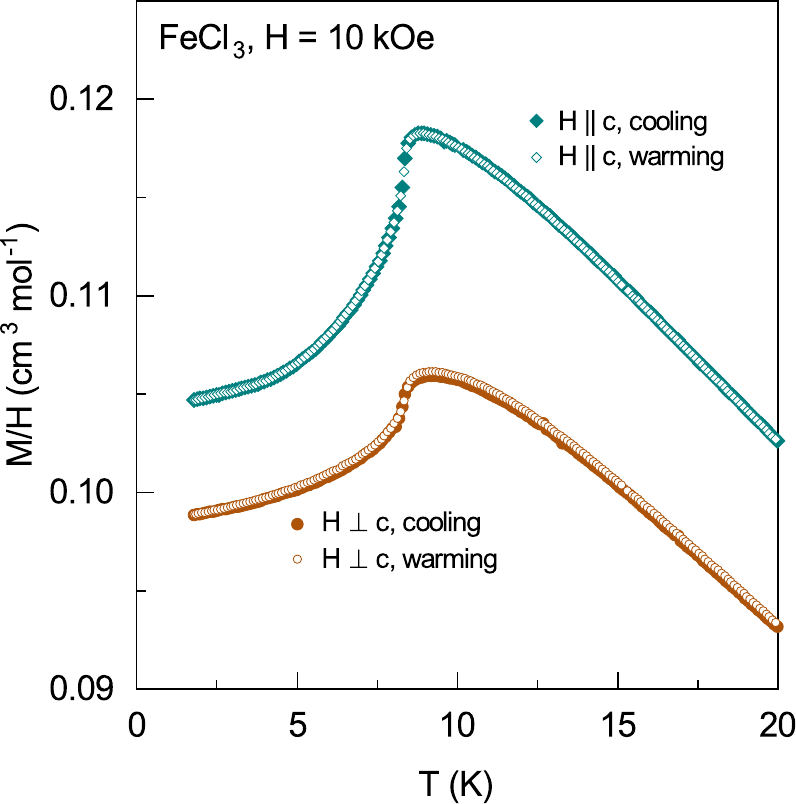}
      \caption{\label{mag} Temperature dependence of the magnetic susceptibility ($M/H$) measured with a field of 10\,kOe applied in the plane ($\perp$c) and out of the plane ($\parallel c$) on cooling and subsequent warming. }
  \end{figure}

  \subsection{Magnetic susceptibility}
  Magnetization measurements were performed using a Quantum Design MPMS3 dc-SQUID magnetometer. Samples were protected from air exposure by loading them into silica tube sample holders that were flame-sealed on one end and sealed with Dow-Corning high vacuum grease at the other. These sample holders were loaded and placed into plastic drinking straws in the glovebox, then quickly transported to the magnetometer and inserted for measurements to minimize reaction with moisture in air.
  
  The temperature dependence of the magnetic susceptibility measured near the magnetic ordering temperature is shown in Fig.~\ref{mag}.  The slight difference between the data measured in field perpendicular or parallel with the $c$ axis might arise from the development of a small orbital moment due to covalence between Fe$^{3+}$ and Cl$^{-}$ ions. Data at higher temperatures are shown in the main text. The observed behavior is generally consistent with previous reports for FeCl$_3$ \cite{starr_magnetic_1940, jones_low_1969s, johnson_field_1981s}. Upon cooling, the susceptibility reaches a relatively broad maximum near 9\,K, with a sharp drop near 8.5\,K. At temperatures above $\sim 100$ K, the Curie-Weiss law is obeyed, with a fitted effective moment of 5.84\,$\mu_B$ per Fe and Weiss temperature of $-17$\,K.
  \begin{figure}[h!]
      \includegraphics[width=0.35\textwidth]{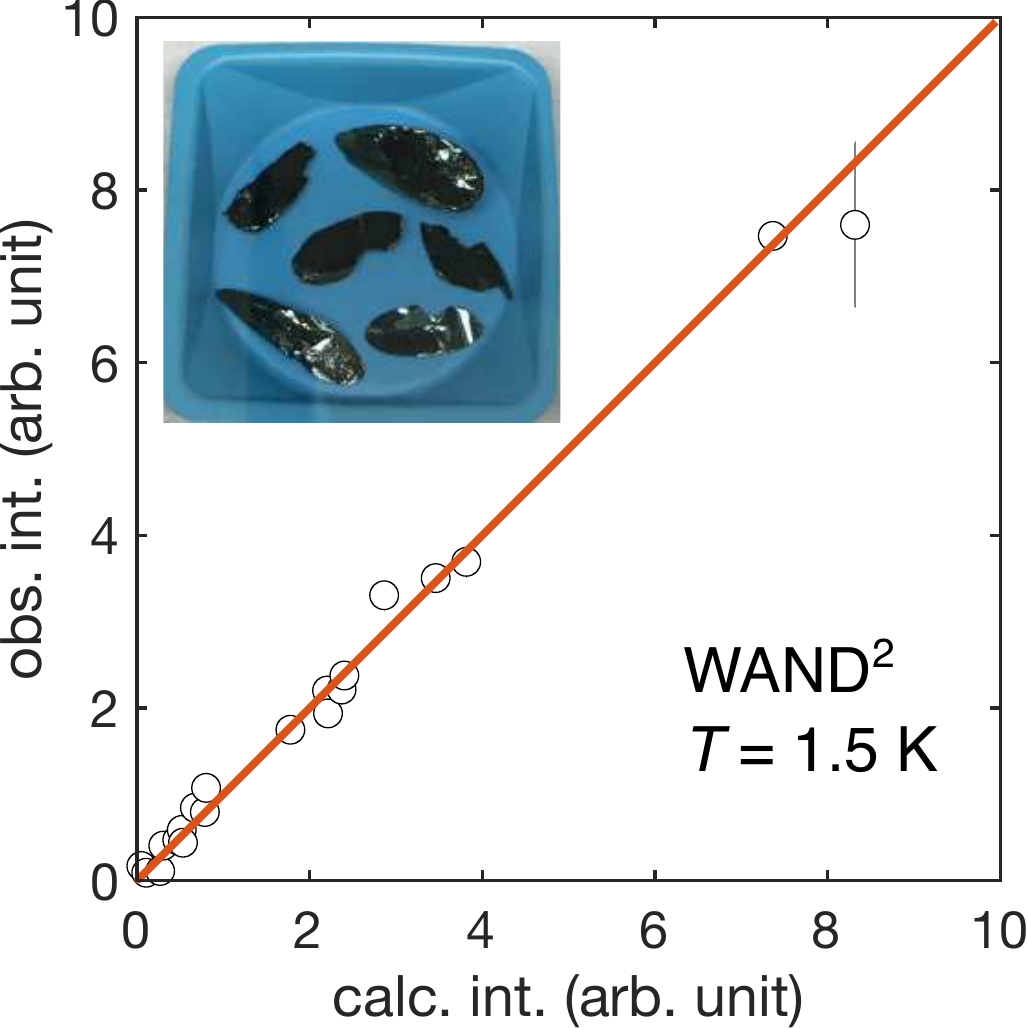}
      \caption{Comparison between the calculated and observed intensities of the nuclear Bragg peaks collected at $T = 1.5$ K. Inset is a picture of the FeCl$_3$ crystals with a typical size of $\sim2.5$~cm in length and $\sim0.2$~cm in thickness. The weigh boat containing the crystals is 9 cm on a side and the photo was taken through the glove box window. }
      \label{fig:wand}
  \end{figure}

  \subsection{Neutron diffraction experiment}
  Neutron diffraction experiments on a FeCl$_3$ crystal were performed on the WAND$^2$ diffractometer~\cite{frontzek_wand_2018} at the High Flux Isotope Reactor (HFIR), Oak Ridge National Laboratory (ORNL). A crystal of $\sim0.15$~g was aligned with the $c$ axis vertical in a sealed aluminum can with 1 atm of He gas to prevent reaction with moisture in air. A cryomagnet was employed to reach the base temperature of 1.5~K. Altogether 46 nuclear reflections, of which 20 are independent, were collected at 1.5 K. Structural refinement was performed using the Fullprof software~\cite{rodriguez_recent_1993}.
  
  Weak stacking faults are observed in our FeCl$_3$ sample as elongated Bragg peaks along the $c$ direction.  Throughout the investigated temperature range, the nuclear Bragg peaks can be indexed by the $R\overline{3}$ space group. Figure~\ref{fig:wand} compares the calculated and observed intensities of the nuclear Bragg peaks collected on WAND$^2$. The goodness-of-fit factors are $R_F = 5.3~\%$ and $R_{F2} = 7.2~\%$. The refined atomic positions are [0,0,0.317(7)] for Fe and [0.357(3), 0, 0.071(8)] for Cl.
  
  \subsection{Diffuse neutron scattering experiment}
  Diffuse neutron scattering experiments were performed on the CORELLI elastic diffuse scattering spectrometer at the Spallation Neutron Source (SNS), ORNL~\cite{ye_implementation_2018}. A crystal of $\sim0.4$~g was aligned with the $c$ axis vertical in a sealed aluminum can with 1 atm of He gas to prevent reaction with moisture in air. A closed cycle refrigerator (CCR) was employed to reach temperatures down to 5 K. Data were acquired by rotating the sample in 2$^{\circ}$ steps, covering a total range of 360$^{\circ}$. Data reduction and projection were performed using the MANTID package~\cite{arnold_mantid_2014}. For the combined simplex/simulated annealing fits, the diffuse scattering data in the range of $h=[-0.5,0.5]$, $k'=[-0.4,0.4]$, and $l=[-2,0]$ were binned in steps of 0.05, 0.04, and 0.1 (r.l.u.), respectively, leading to a total of 7983 data points after excluding the unmeasured positions. Preliminary diffuse neutron scattering experiments were also performed on the HB2A POWDER diffractometer at the HFIR, ORNL. The results (not shown) are consistent with the data collected on CORELLI.
  
  \subsection{Inelastic neutron scattering experiment.}
  INS experiments were performed on the SEQUOIA spectrometer at the SNS. A FeCl$_3$ crystal of $\sim 0.7$~g was aligned with the $(h,0,l)$ plane horizontal in a CCR. The sample was sealed in an aluminum can with 1 atm of He gas to prevent reaction with moisture in air. An incident neutron energy of $E_i$ = 8~meV was used in the high resolution mode with a Fermi chopper frequency of 120~Hz. Data were acquired by rotating the sample in 1$^{\circ}$ steps, covering a total range of 180$^{\circ}$. Data reduction and projection were performed using the MANTID~\cite{arnold_mantid_2014} and HORACE~\cite{ewings_horace_2016} packages. Linear spin wave calculations were performed using the SpinW program~\cite{toth_linear_2015}. Preliminary INS experiments were also performed on the ARCS spectrometer at the SNS. The results (not shown) are consistent with the data collected on SEQUOIA. 
  
  \subsection{Weak stacking faults}
\begin{figure}[h!]
      \includegraphics[width=0.4\textwidth]{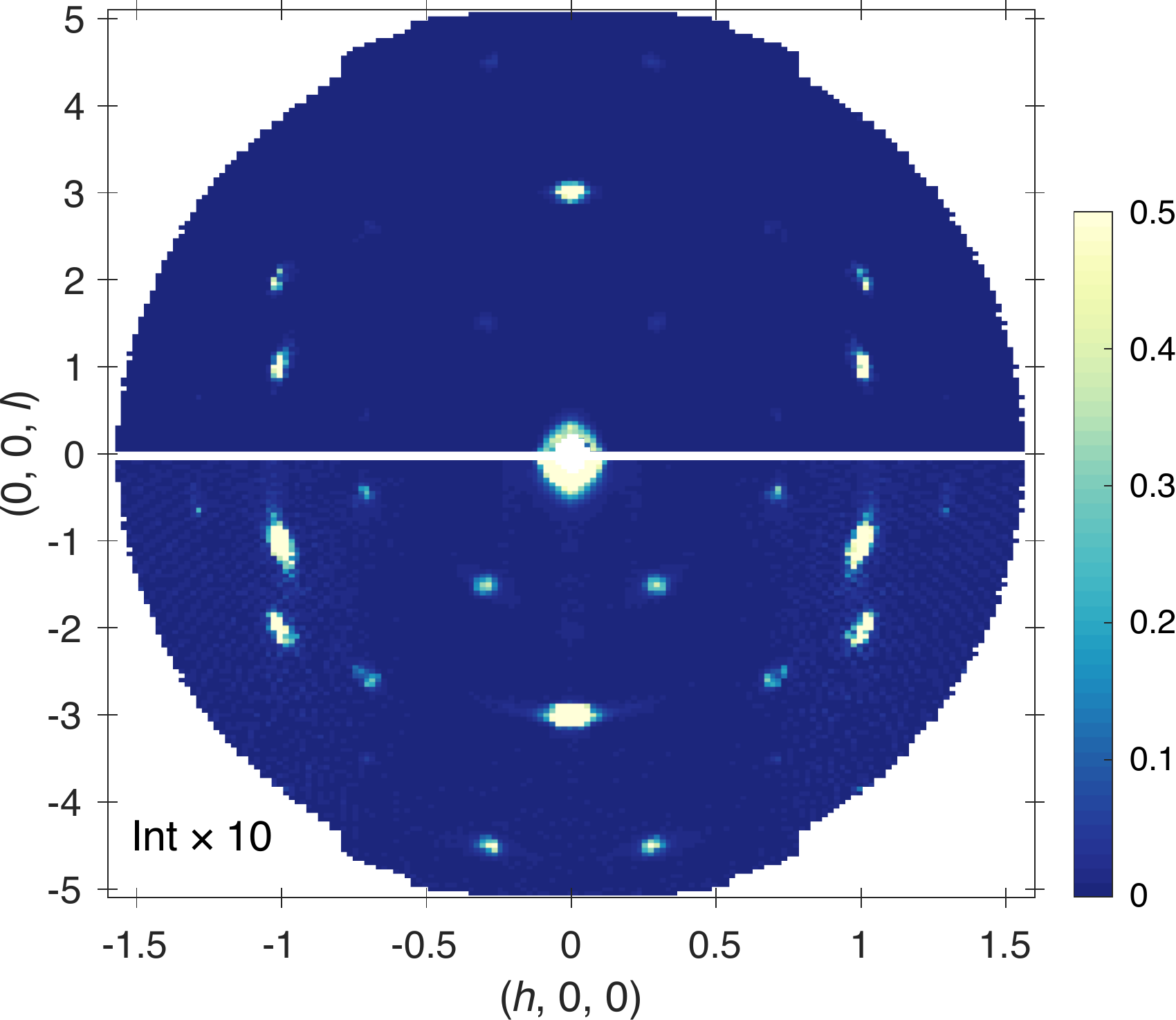}
      \caption{Elastic slice of the SEQUOIA data in the $k=0$ plane collected at $T=4$~K with $E_i = 8$~meV. The integration ranges are $k=[-0.05,0.05]$ r.l.u. and $E=[-0.25,0.25]$ meV. Data have been symmetrized according to the $R\overline{3}$ space group symmetry. Intensity in the lower half panel is multiplied by 10 times to expose the weak streaks underneath the nuclear Bragg peaks.}
      \label{fig:elastic}
      \end{figure}

  Due to the weak bonding between the adjacent layers, the van der Waals materials often contain stacking faults~\cite{johnson_monoclinic_2015, kong_crystal_2019}. The elastic slice of our SEQUOIA data in the $k=0$ plane is shown in Fig.~\ref{fig:elastic}. In the lower half panel, intensity is multiplied by 10 times to expose the weak streaks underneath the nuclear Bragg peaks, which reveals the presence of weak stacking faults in our FeCl$_3$ sample. However, magnetic Bragg peaks, \textit{e.g.} in the $l = -1.5$ plane, are relatively sharp along all directions, indicating a marginal effect of stacking faults on the magnetic correlations.
  
  \subsection{Diffuse patterns in a wide range}
\begin{figure}[h!]
      \includegraphics[width=0.9\textwidth]{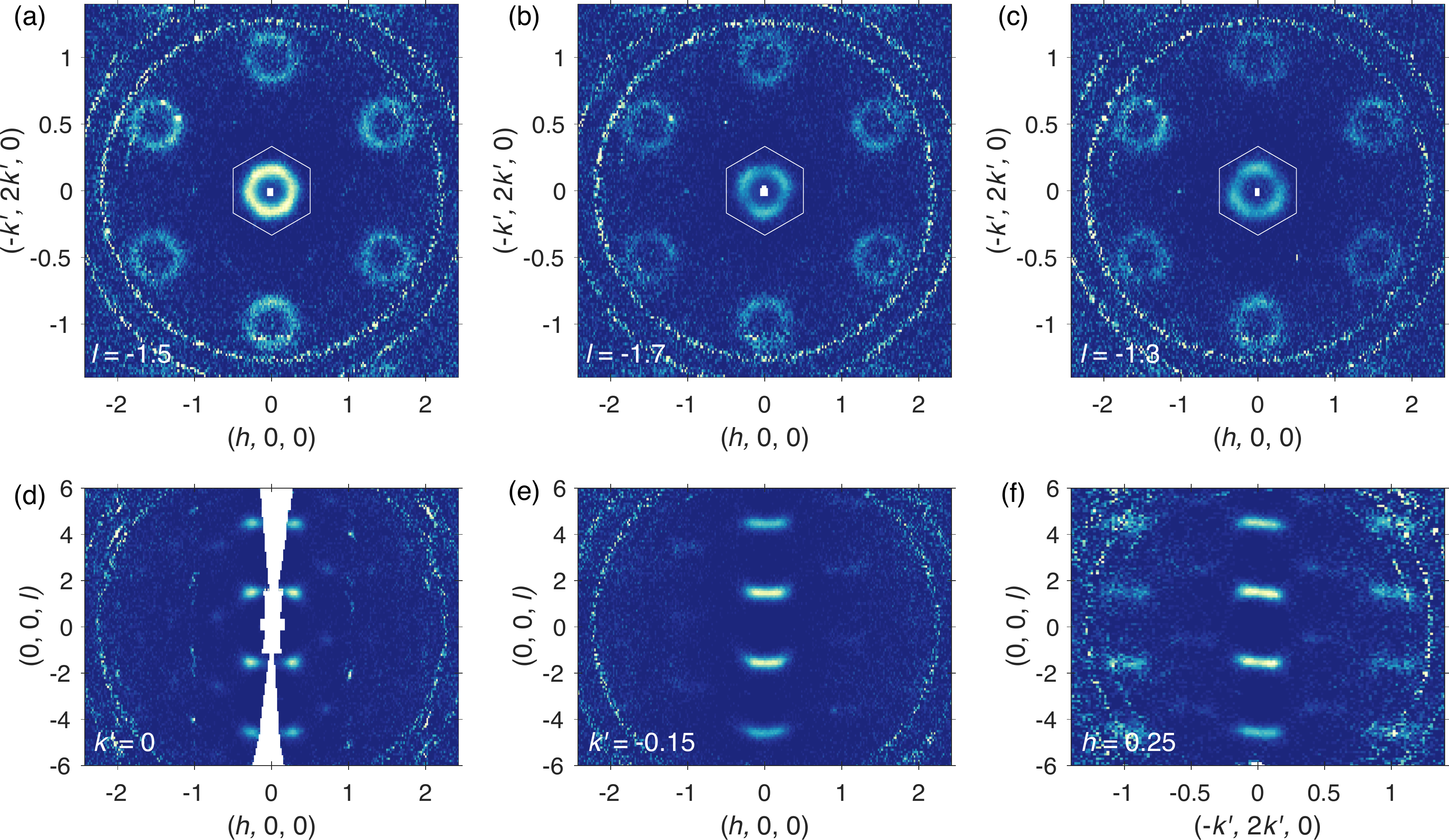}
      \caption{Diffuse neutron scattering intensity measured at 10 K in a wide range. Data in the $l = -1.5$, $-1.7$, and $-1.3$ planes are shown in panels (a-c), respectively. Data in the $k'=0$, $k'=-0.15$, and $h = 0.25$ planes are shown in panels (d-f), respectively.}
      \label{fig:largeQ}
      \end{figure}

  Figure~\ref{fig:largeQ} presents the diffuse scattering patterns collected at 10 K with CORELLI in a wider range compared to those in Fig.~3(a-f) of the main text.

  \subsection{The self-consistent Gaussian approximation (SCGA)}
  
  Neutron scattering patterns for the short-range spin correlations were calculated using the SCGA method~\cite{conlon_absent_2010, paddison_scattering_2020, gao_suppressed_2021}. The Hamiltonian of a Heisenberg model with general interactions (isotropic and anisotropic) can be written as~\cite{enjalran_theory_2004, toth_linear_2015s}
  \begin{align}
  \mathcal{H} = \frac{1}{2}\sum_{\substack{m\mu\alpha \\ n\nu\beta}}\mathcal{J}_{m\mu, n\nu}^{\alpha\beta} S_{m\mu}^\alpha S_{n\nu}^{\beta}\ \textrm{,}
  \end{align}
  where $S^{\alpha}_{m\mu}$ is the spin component along $\alpha=x,y,z$ at sublattice $\mu$ in unit cell $m$. With Fourier transform, we have
  \begin{align}
      S^{\alpha}_{\mu}(\bm{q}) = \frac{1}{N}\sum_m \exp{(-i\bm{q}\cdot \bm{r}_{m\mu})}S^\alpha_{m\mu}\ \textrm{,}
  \end{align}
  where $\bm{r}_{m\mu}$ is the position of the spin $\bm{S}_{m\mu}$ and $N$ is the number of unit cells. The interaction matrix becomes
  \begin{align}
      \mathcal{J}_{\mu\nu}^{\alpha\beta} (\bm{q})= \sum_{m,n}\exp[-i\bm{q}\cdot (\bm{r}_{m\mu}-\bm{r}_{n\nu})]\mathcal{J}_{m\mu,n\nu}^{\alpha\beta}\ \textrm{.}
  \end{align}
  At each $\bm{q}$ position, $J_{\mu\nu}^{\alpha\beta}(\bm{q})$ forms a $3M\times3M$ matrix with $M$ denoting the number of sublattices, and its eigenvalues and eigenvectors can be denoted as $\omega_\rho(\bm{q})$ and $\bm{U}_\rho(\bm{q})$, respectively. Under the Gaussian approximation, the spin-spin correlations can be calculated by
  \begin{align}
      \langle S_\mu^{\alpha}(-\bm{q}) S_\nu^{\beta}(\bm{q})\rangle =\sum_{\rho} \frac{U_\rho^{\mu\alpha}(\bm{q}) U_\rho^{\nu\beta}(\bm{q})^*}{\lambda+\beta\omega_\rho(\bm{q})} \textrm{,}
  \end{align}
  where $\beta = 1/k_\textrm{B}T$ and $\lambda$ is determined self-consistently by the spin length constraint
  \begin{align}
   \frac{1}{MN}\sum_{\bm{q}\in \textrm{BZ},\rho}\frac{1}{\lambda+\beta\omega_\rho(\bm{q})} = 1 \textrm{.}
  \end{align}
  Finally, the neutron scattering cross section can be calculated as
  \begin{align}
  \frac{d\sigma(\bm{q})}{d\Omega} = C[f(|\bm{q}|)]^2\sum_{\mu\alpha,\nu\beta}(\delta^{\alpha\beta} - \frac{q^\alpha q^\beta}{|\bm{q}|^2})\langle S_\mu^{\alpha}(-\bm{q}) S_\nu^{\beta}(\bm{q})\rangle\ \textrm{,}
  \end{align}
  where $C$ is a constant and $f(\bm{q})$ is the magnetic form factor of the Fe$^{3+}$ ions.
  
  For Heisenberg models, the magnetic susceptibility is isotropic, and the component along the assumed field direction $z$ is calculated as
  \begin{align}
      \chi^z(T) = \frac{1}{NT}\sum_{ij}\langle S_i^z S_j^z\rangle\ \textrm{,}
  \end{align}
  where $\langle S_i^z S_j^z\rangle = \langle S_i^z(0) S_j^z(0)\rangle$ is calculated using Eq.~(4) at $\bm{q}=0$. We assume $S = 5/2$ for the Fe$^{3+}$ ions throughout our calculations.

  \subsection{Diffuse scattering of the $J_1$-$J_2$-$J_{c1}$ minimal model}
  
\begin{figure}[h!]
      \includegraphics[width=0.6\textwidth]{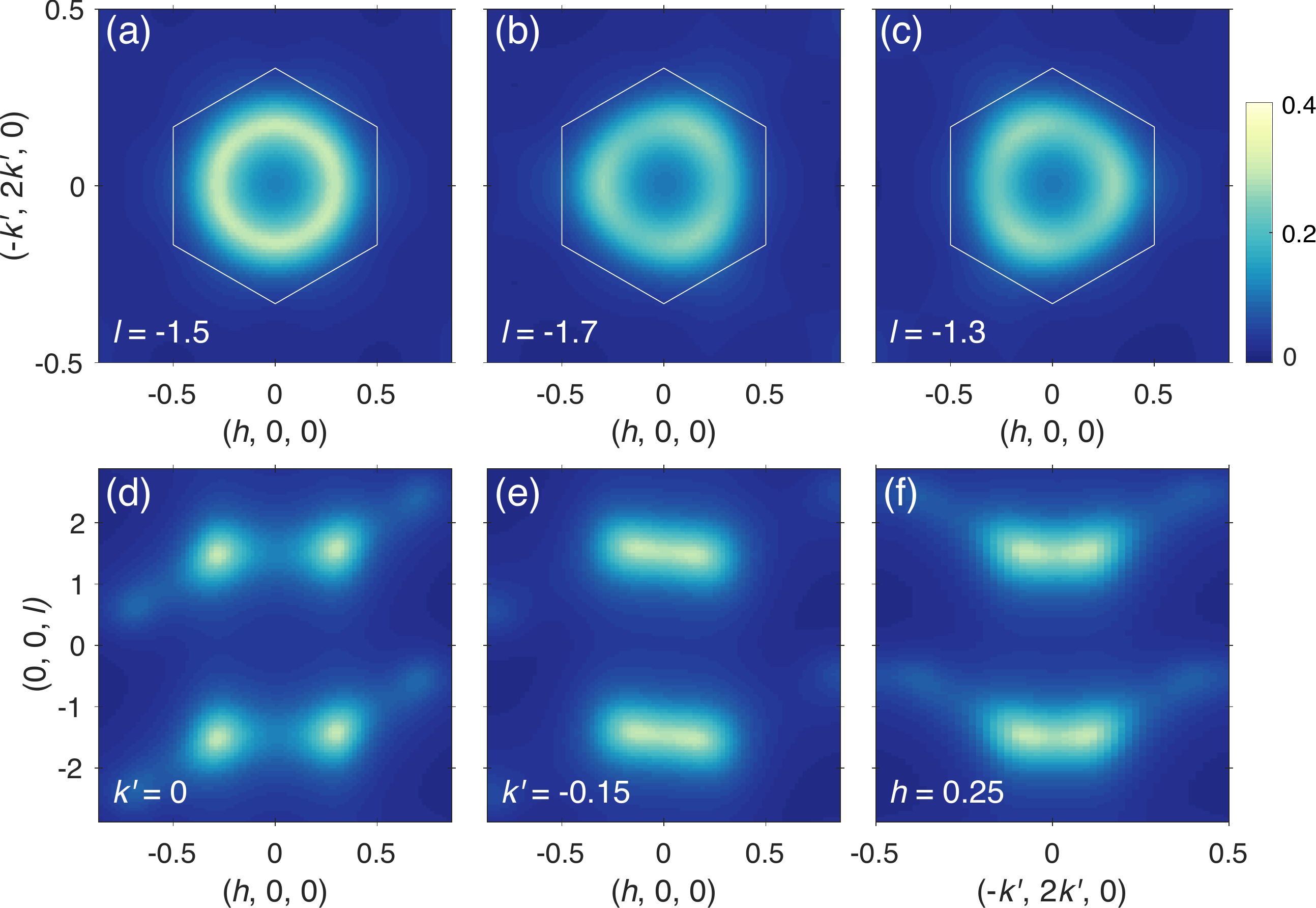}
      \caption{Calculated diffuse scattering patterns for the $J_1$-$J_2$-$J_{c1}$ minimal model at $T=10$~K with $J_1 = -0.3$~meV, $J_2 = 0.075$~meV, and $J_{c1}=0.15$~meV. The same color scheme is used for all the panels.
      \label{fig:minimal}}
      \end{figure}

  Figure~\ref{fig:minimal} presents the diffuse scattering patterns for the $J_1$-$J_2$-$J_{c1}$ minimal model calculated at $T = 10$~K. The calculated intensity in the ($h,k,\overline{1.7}$) and ($h,k,\overline{1.3}$) planes exhibits weak three-fold patterns that are rotated by $\sim30^{\circ}$ compared to the experimental data as presented in the main text. Such a deviation indicates the existence of further perturbations to the minimal model.

  \begin{figure}[t!]
      \includegraphics[width=0.6\textwidth]{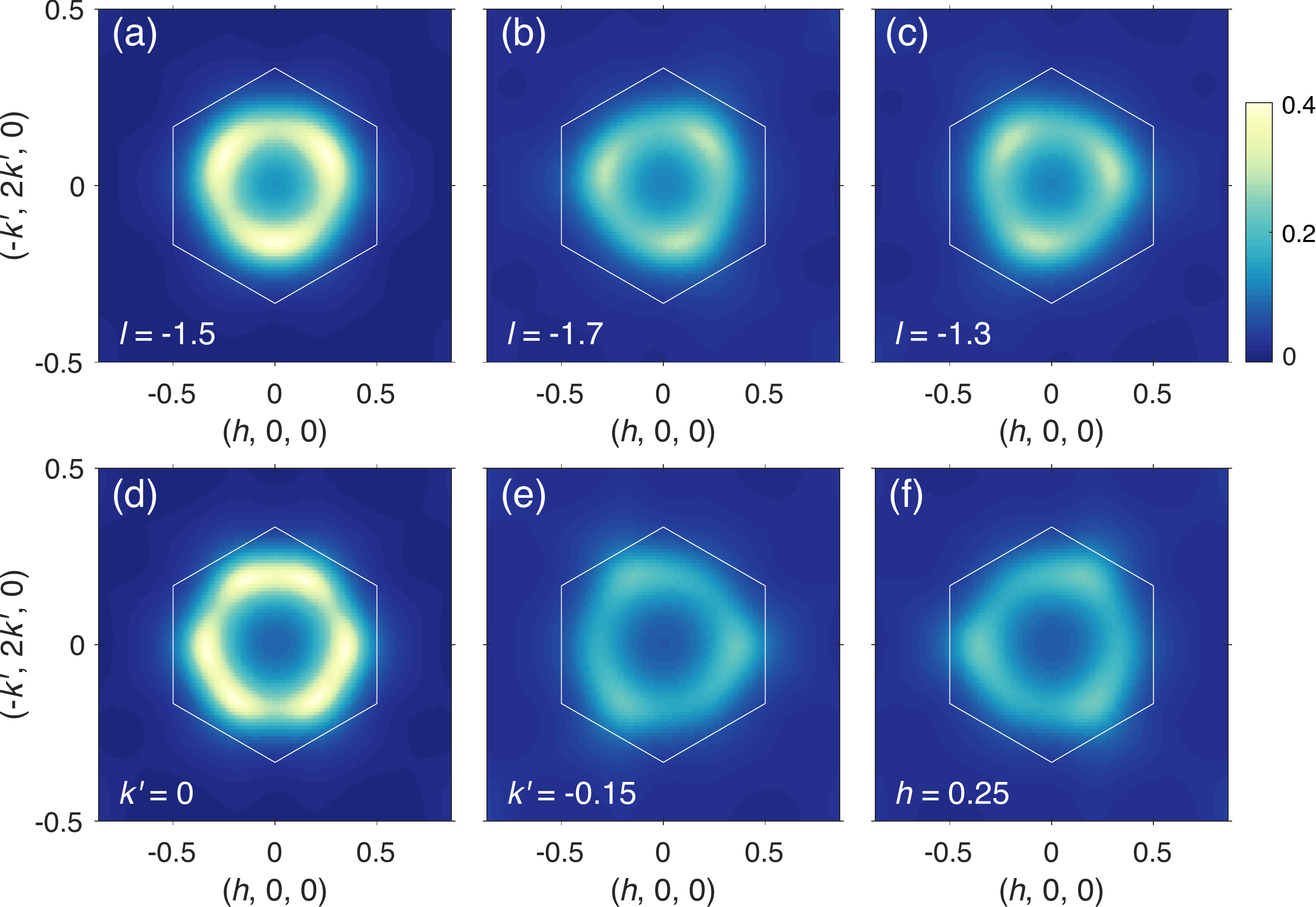}
      \caption{Calculated diffuse scattering patterns for the $J_1$-$J_2$-$J_{c1}$ minimal model being perturbed by the Kitaev couplings over the $J_1$ bonds with (a-c) $\phi = \pi/16$ and (d-f) $\phi =- \pi/16$. The calculation temperature is 10~K.  The same color scheme is used for all the panels.
      \label{fig:kitaev}}
      \end{figure}
  
  \begin{figure}[h!]
      \includegraphics[width=0.6\textwidth]{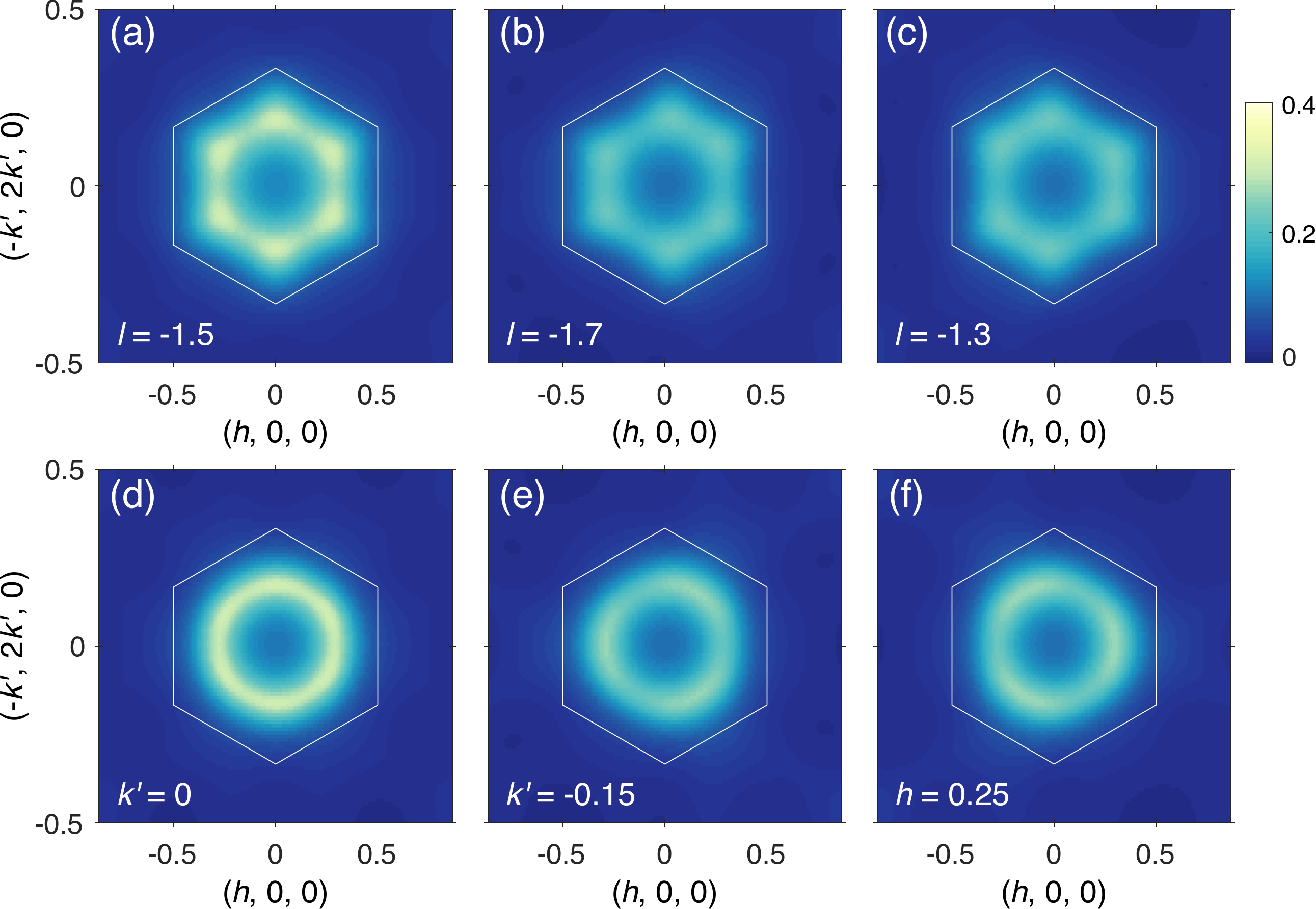}
      \caption{Calculated diffuse scattering patterns for the $J_1$-$J_2$-$J_{c1}$ minimal model being perturbed by the Dzyaoshinskii-Moriya interactions over the $J_2$ bonds with the DM vectors along the $c$ axis (a-c) and within the $ab$ plane (d-f). The calculation temperature is 10~K. The same color scheme is used for all the panels.
      \label{fig:dmi}}
      \end{figure}

  \subsection{Perturbations from anisotropic interactions}
  
  Besides the perturbations from the isotropic further-neighbor interactions, we also explore the perturbations from the anisotropic exchange interactions, which include the Kitaev interactions over the $J_1$ bonds and the Dzyaloshinskii-Moriya (DM) interactions on the $J_2$ bonds, both being allowed by symmetry. 
  
  Following Ref.~\cite{rau_generic_2014}, the isotropic and Kitaev couplings over the $J_1$ bonds are parameterized as $J_1 = J\cos\phi$ and $K_1=J\sin\phi$, respectively, where $J = -0.30$~meV as discussed for the $J_1$-$J_2$-$J_{c1}$ minimal model. Figure~\ref{fig:kitaev} summarizes the diffuse patterns for a perturbed minimal model with $\phi= \pi/16$ and $-\pi/16$, which corresponds to the case of $K_1 = 0.2J_1$ and $-0.2J_1$, respectively. In both cases, the scattering intensity along the spiral ring in the $(h,k,\overline{1.5})$ plane is strongly modulated, and the three-fold patterns in the $(h,k,\overline{1.7})$ and $(h,k,\overline{1.3})$ planes are not  compatible with the experimental data. 
  
  For the DM interactions, we assume the DM vectors $\bm{D}$ to be perpendicular to the $J_2$ bonds and consider two cases with the DM vectors pointing along the $c$ axis~\cite{chen_topological_2018} or being parallel with the $ab$ plane. Figure~\ref{fig:dmi} summarizes the calculated diffuse patterns for the perturbed minimal model with a DM strength of $0.2|J_1|$. When the DM vectors are pointing along the $c$ axis (Figs.~\ref{fig:dmi}(a-c)), the diffuse patterns in all the $(h,k,\overline{1.7})$, $(h,k,\overline{1.5})$, and $(h,k,\overline{1.3})$ planes exhibit a strong hexagonal distortion that is inconsistent with the experimental data. When the DM vectors are parallel with the $ab$ plane, the diffuse patterns are similar to those of the unperturbed $J_1$-$J_2$-$J_{c1}$ model. Thus we conclude that anisotropic interactions are not the main perturbations in FeCl$_3$. 
  
  \begin{figure*}[t!]
      \includegraphics[width=0.8\textwidth]{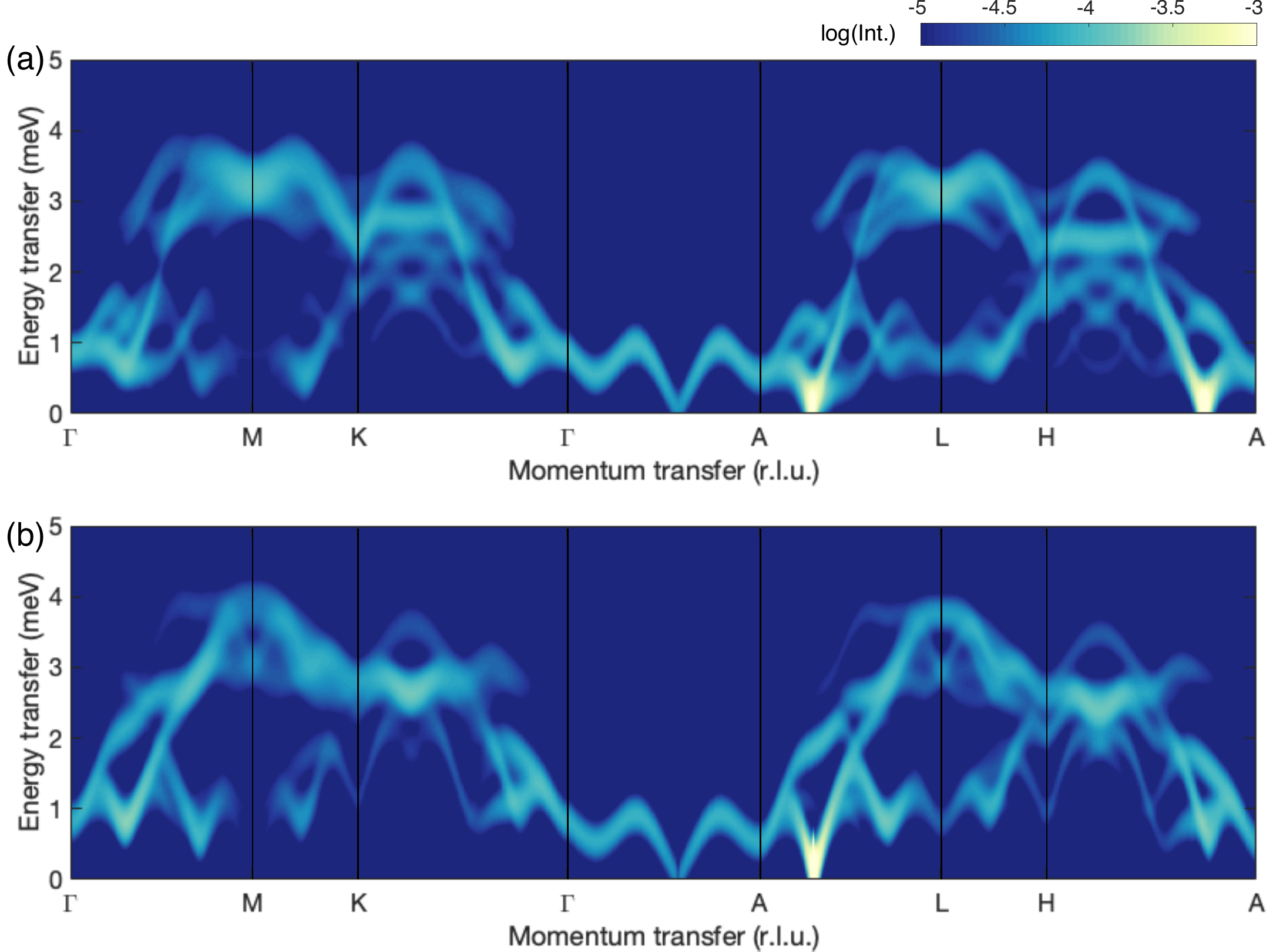}
      \caption{Calculated INS spectra for the (a) $J_1$-$J_4$-$J_{c1}$-$J_{s1}$ and (b) $J_1$-$J_5$-$J_{c1}$-$J_{s1}$ models with coupling strengths shown in the text. For both calculations, a weak easy-plane single ion anisotropy of 1~$\mu$eV is applied to stabilize the helical ground state.
      \label{fig:other}}
      \end{figure*}
  \begin{figure}[b!]
      \includegraphics[width=0.8\textwidth]{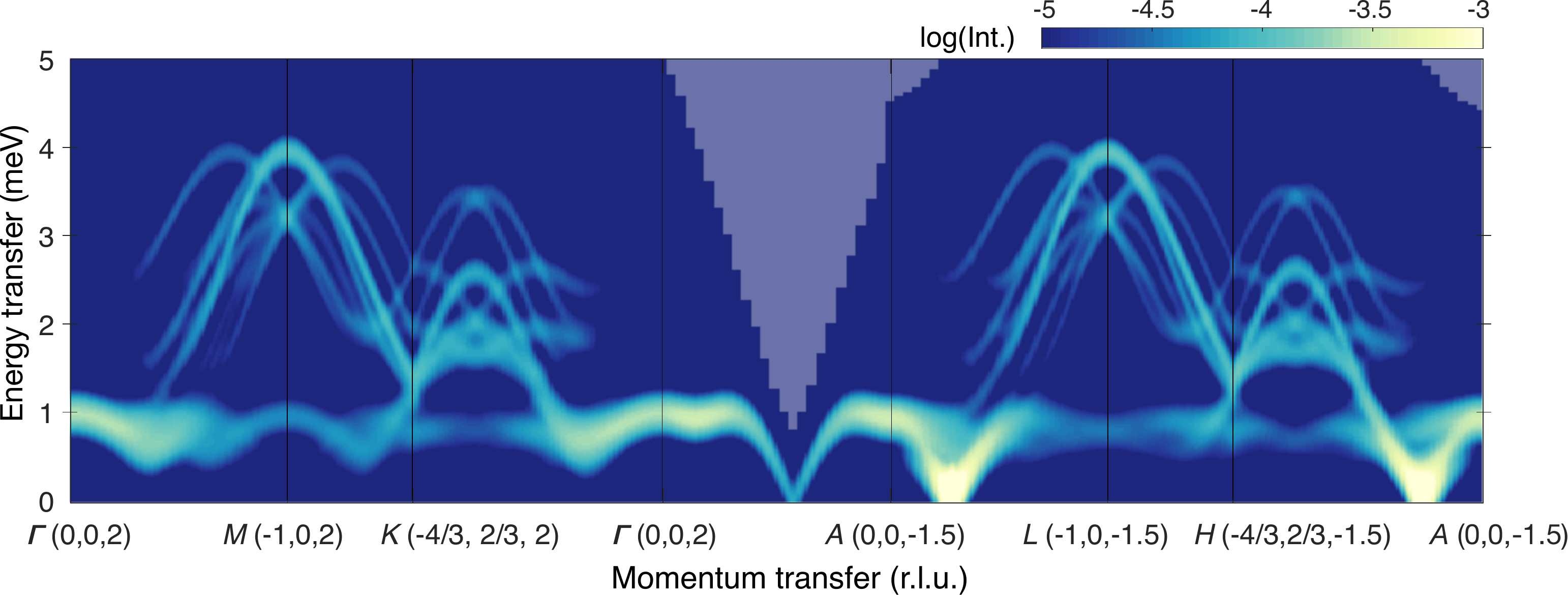}
      \caption{Calculated INS spectra for the $J_{123}$-$J_{c123}$ model with coupling strengths fitted from the diffuse scattering data.
      \label{fig:insdiff}}
      \end{figure}
  
  \subsection{$J_1$-$J_4$ and $J_1$-$J_5$ competitions}
  Besides the $J_1$-$J_2$ model, ring-like scattering may also arise from the competition between $J_1$ with the fourth-neighbor coupling $J_4$ or the fifth-neighbor coupling $J_5$ in the honeycomb layer. Therefore, INS spectra are also calculated for the $J_1$-$J_4$-$J_{c1}$-$J_{s1}$ and $J_1$-$J_5$-$J_{c1}$-$J_{s1}$ models, where $J_{c1}$ and $J_{s1}$ as defined in the main text are included as the minimal interlayer couplings to reproduce the magnon dispersion along the $\Gamma$-$A$ line. We verify that the shape of the magnon dispersions in the honeycomb layer is not modulated by the $J_{c1}$ and $J_{s1}$ couplings. Figure~\ref{fig:other} presents the calculated INS spectra for the $J_1$-$J_4$-$J_{c1}$-$J_{s1}$ and $J_1$-$J_5$-$J_{c1}$-$J_{s2}$ models. For the $J_1$-$J_4$-$J_{c1}$-$J_{s1}$ model, the ratio of $J_4$/$J_1$ is fixed at 0.35 so that the propagation vector $\bm{q}=(0.168, 0.168, 1.5)$ of the ground state lies on the spiral ring. The strengths of $J_1$, $J_{c1}$, and $J_{c2}$ are $-0.27$, $0.039$, and $-0.075$~meV, respectively. For the $J_1$-$J_5$-$J_{c1}$-$J_{s1}$ model, the ratio of $J_5$/$J_1$ is fixed at 0.6 so that the ground state propagation vector $\bm{q}=(0.296, 0, 1.5)$ is close to the spiral ring. The strengths of $J_1$, $J_{s1}$, and $J_{s2}$ couplings are $-0.225$, 0.032, and $-0.062$~meV, respectively. The calculated INS spectra deviate from our experimental results, thus confirming the spiral ring in FeCl$_3$ originates from the $J_1$-$J_2$ competition instead of the $J_1$-$J_4$ or $J_1$-$J_5$ competition.
  
  \subsection{INS spectra of the $J_{123}$-$J_{c123}$ model}
  Figure~\ref{fig:insdiff} shows the INS spectra calculated for the $J_{123}$-$J_{c123}$ model with coupling strengths fitted from the diffuse scattering data as discussed in the main text. Although the dispersion in the honeycomb layer is similar to the experimental data, the gull wing-shaped dispersion along the $\Gamma$-$A$ high symmetry line is not reproduced. 
  
  \begin{figure}[t!]
      \includegraphics[width=0.6\textwidth]{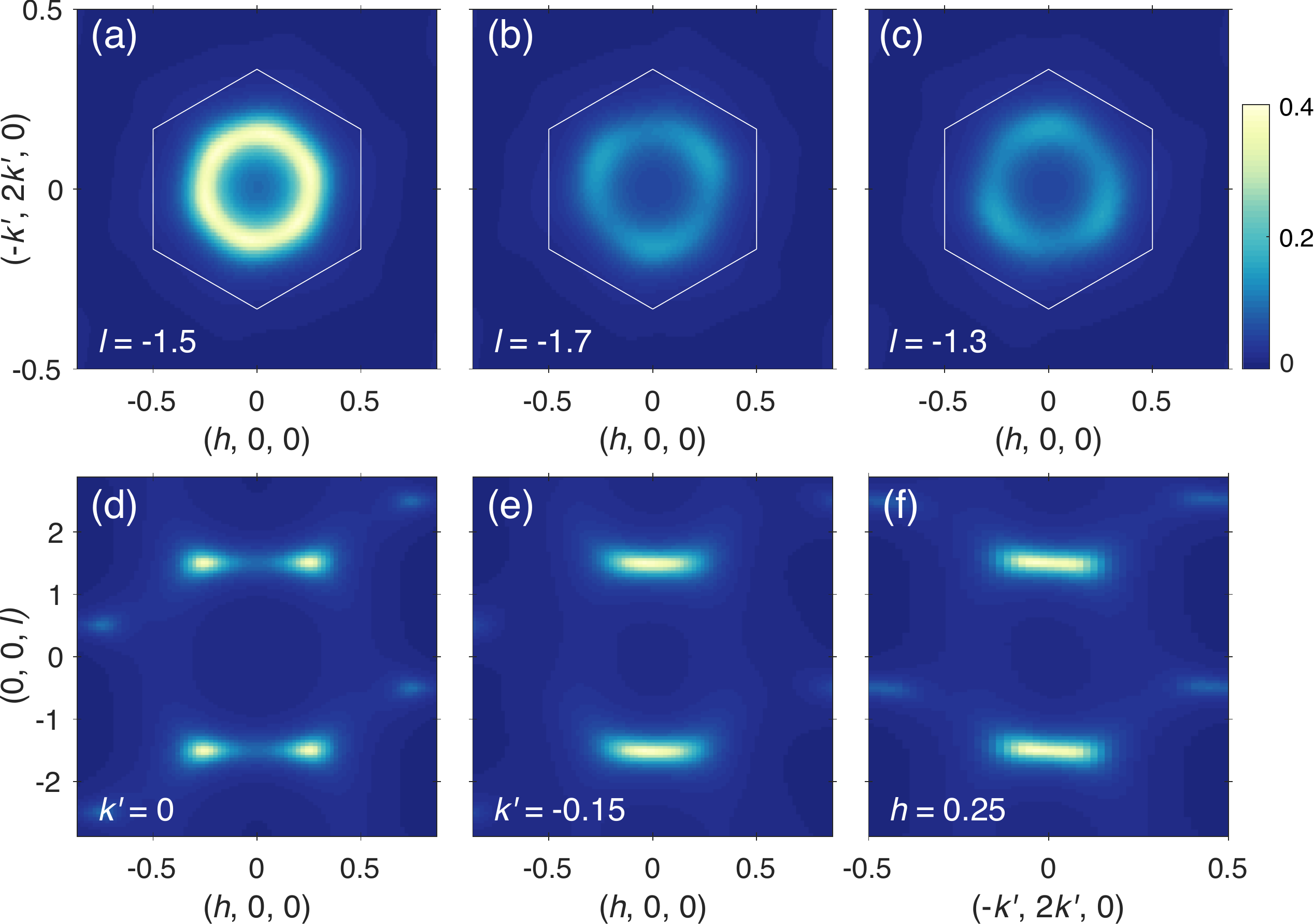}
      \caption{Calculated diffuse patterns for the $J_{123}$-$J_{c123}$-$J_{s123}$ model with coupling strengths fitted from the INS data. Slices are the same as those in Fig.~3 of the main text.
      \label{fig:diffins}}
      \end{figure}

  \subsection{Diffuse patterns of the $J_{123}$-$J_{c123}$-$J_{s123}$ model}
  Figure~\ref{fig:diffins} shows the diffuse patterns calculated for the $J_{123}$-$J_{c123}$-$J_{s123}$ model with coupling strengths fitted from the INS data as discussed in the main text. Similar to the $J_{123}$-$J_{c123}$ model, the calculated patterns reproduce the experimental data, including the three-fold patterns in the $(h,k,\overline{1.7})$ and $(h,k,\overline{1.3})$ planes.
  
  \subsection{Magnetic frustration in the honeycomb-lattice compounds}
  Magnetic frustration characterized by the ratio of $|J_2/J_1|$ is generally weak in honeycomb-lattice compounds. Table~\ref{tab:honey} lists the strengths of the $J_1$ and $J_2$ interactions together with the ratio of $|J_2/J_1|$ and the magnetic ground state. Many of the compounds are free from frustration, as both $J_1$ and $J_2$ are FM. Among the compounds with non-zero frustration, the ratio of $|J_2/J_1|$ is less than 1/6. One exception is ZnMnO$_3$, where a relatively high ratio of $|J_2/J_1|=0.44$ has been proposed based on the magnetic susceptibility measurements. Further studies with neutron scattering will be required to verify the proposed ratio.
  \begin{table}[b!]
      \caption{The strengths of the $J_1$ and $J_2$ interactions in unit of meV for some representative honeycomb-lattice compounds. The ratio of $|J_2/J_1|$ and the magnetic ground state are shown in the last two columns. The exchange parameters obtained from the diffuse scattering on FeCl$_3$ are also included for comparison.}
      \label{tab:honey}
      \centering
      \begin{tabular}{l@{\hskip 0.3in}c@{\hskip 0.3in}c@{\hskip 0.3in}c@{\hskip 0.3in}l}
      \toprule
      Compound [Ref] & \, $J_1$  &\, $J_2$ & $|J_2/J_1|$\, &\, Ground state\\
          \hline
      ZnMnO$_3$~\cite{haraguchi_frustrated_2019s}   & $J_1<0$  & $J_2>0$ & 0.44 & N\'eel, $q = (1/2, 1/2, 0)$\\
      \textbf{FeCl$_3$~(this work)} & $\bm{-0.25}$ & \textbf{0.09} & \textbf{0.36} & \textbf{helical,} $\bm{q = (\frac{4}{15},\frac{1}{15},\frac{3}{2})$}\\
      Bi$_3$Mn$_4$O$_{12}$(NO$_3$)~\cite{matsuda_frustrated_2019} & 3.3 & 0.46 & 0.14 & disorder \\
      MgMnO$_3$~\cite{haraguchi_frustrated_2019s}   & $J_1>0$  & $J_2>0$ & 0.13 & N\'eel, $q = 0$\\
      YbBr$_3$~\cite{wessler_observation_2020}   & $0.69$  & $0.09$ & 0.13 & N\'eel, $q = 0$\\
      CrGeTe$_3$~\cite{zhu_topo_2021} & $-2.73$ & $-0.33$ & 0.12 & FM, $q=0$\\
      CrSiTe$_3$~\cite{zhu_topo_2021} & $-1.49$ & $-0.15$ & 0.10 & FM, $q=0$\\
      CrI$_3$~\cite{chen_topological_2018} & $-2.01$ & $-0.16$ & 0.08 & FM, $q =0$\\
      CrBr$_3$~\cite{cai_topological_2021} & $-1.36$ & $-0.06$ & 0.04 & FM, $q = 0$\\
      CrCl$_3$~\cite{chen_massless_2021} & $-0.95$ & $-0.024$ & 0.03 & layer AF, $q = (0,0,3/2)$\\
      \botrule
      \end{tabular}
      \end{table}

%
 
\end{document}